\documentclass[10pt,conference]{IEEEtran}
\IEEEoverridecommandlockouts
\usepackage{cite}
\usepackage{amsmath,amssymb,amsfonts}
\usepackage{algorithmic}
\usepackage{graphicx}
\usepackage{textcomp}
\usepackage{xcolor, colortbl}
\usepackage{xspace}
\usepackage{multirow}
\usepackage{caption}
\usepackage{subcaption}
\usepackage{footnote}
\usepackage{url}
\usepackage{tcolorbox}
\usepackage{booktabs}
\usepackage{enumitem}
\usepackage{fixltx2e}
\usepackage{multicol}
\usepackage{balance}
\usepackage{listings}
\usepackage{todonotes}
\usepackage[normalem]{ulem}

\definecolor{Gray}{gray}{0.9}
\definecolor{codegreen}{rgb}{0,0.6,0}
\definecolor{codegray}{rgb}{0.73,0.38,0.06}
\definecolor{codepurple}{rgb}{0.70,0.27,0}
\definecolor{codemagenta}{rgb}{0.74,0.09,0.42}
\definecolor{codeoutput}{rgb}{0.5,0,0}
\definecolor{backcolour}{rgb}{0.96,0.96,0.96}

\lstdefinelanguage{docker}{
  keywords={FROM, RUN, COPY, ADD, ENTRYPOINT, CMD,  ENV, ARG, WORKDIR, EXPOSE, LABEL, USER, VOLUME, STOPSIGNAL, ONBUILD, MAINTAINER},
  keywordstyle=\color{blue}\bfseries,
  identifierstyle=\color{black},
  sensitive=false,
  comment=[l]{\#},
  commentstyle=\color{purple}\ttfamily,
  stringstyle=\color{red}\ttfamily,
  morestring=[b]',
  morestring=[b]"
}

\definecolor{eclipseStrings}{RGB}{42,0.0,255}
\definecolor{eclipseKeywords}{RGB}{127,0,85}
\colorlet{numb}{magenta!60!black}
\lstdefinelanguage{json}{
    numberstyle=\tiny\color{codegray},
    commentstyle=\color{eclipseStrings}, 
    stringstyle=\color{eclipseKeywords}, 
    string=[s]{"}{"},
    comment=[l]{:\ "},
    morecomment=[l]{:"},
    literate=
        *{0}{{{\color{numb}0}}}{1}
         {1}{{{\color{numb}1}}}{1}
         {2}{{{\color{numb}2}}}{1}
         {3}{{{\color{numb}3}}}{1}
         {4}{{{\color{numb}4}}}{1}
         {5}{{{\color{numb}5}}}{1}
         {6}{{{\color{numb}6}}}{1}
         {7}{{{\color{numb}7}}}{1}
         {8}{{{\color{numb}8}}}{1}
         {9}{{{\color{numb}9}}}{1}
}

\lstdefinestyle{mystyle}{
    backgroundcolor=\color{white},   
    commentstyle=\color{codegreen},
    keywordstyle=\color{codepurple},
    numberstyle=\tiny\color{codegray},
    stringstyle=\color{codemagenta},
    language=docker,
    breakatwhitespace=false,         
    breaklines=true,                 
    keepspaces=true,
    xleftmargin=10pt,
    framexleftmargin=10pt,                 
    numbers=left,                    
    numbersep=5pt,                  
    showspaces=false,                
    showstringspaces=false,
    showtabs=false,                  
    tabsize=2,
    frame=tb,
    framerule=0.5pt,
    basicstyle=\fontsize{5.5}{5.5}\fontfamily{\ttdefault}\selectfont
}
\lstdefinestyle{mystyleresult}{
    backgroundcolor=\color{backcolour},   
    commentstyle=\color{codegreen},
    keywordstyle=\color{codeoutput},
    numberstyle=\tiny\color{codegray},
    stringstyle=\color{red},
    language=Dockerfile,
    breakatwhitespace=false,         
    breaklines=true,                 
    keepspaces=true,                 
    numbers=none,                    
    numbersep=5pt,                  
    showspaces=false,                
    showstringspaces=false,
    showtabs=false,                  
    tabsize=2,
    frame=tb,
    framerule=0pt,
    basicstyle=\color{codeoutput}\fontsize{5.5}{5.5}\fontfamily{\ttdefault}\selectfont
}

\newcommand{\ie}{\emph{i.e.,}\xspace}
\newcommand{\eg}{\emph{e.g.,}\xspace}

\newcommand{\etal}{\emph{et~al.}\xspace}
\newcommand{\secref}[1]{Section~\ref{#1}\xspace}
\newcommand{\figref}[1]{Fig.~\ref{#1}\xspace}

\newcommand{\tabref}[1]{Table~\ref{#1}\xspace}

\newcommand{\RQ}[1]{RQ$_{\textbf{#1}}$\xspace}

\makeatletter
\newcommand\footnoteref[1]{\protected@xdef\@thefnmark{\ref{#1}}\@footnotemark}
\makeatother

\newboolean{showcomments}
\setboolean{showcomments}{true}
\ifthenelse{\boolean{showcomments}}{
  \marginparwidth=1.35cm
  \newcommand{\nb}[2]{\todo[size=\tiny, color=green!40, nolist, fancyline]{\textbf{\textsc{#1}}. #2}}
}{
  \newcommand{\nb}[2]{}
}

\newcommand{\nlRecipe}{HLS\xspace}
\newcommand{\nlRecipes}{HLSes\xspace}
\newcommand{\nlRecipeName}{high-level specification\xspace}
\newcommand{\nlRecipeNames}{high-level specifications\xspace}

\newcommand{\datasetFinetuningTrain}{D\textsubscript{FT-train}\xspace}
\newcommand{\datasetFinetuningEval}{D\textsubscript{FT-eval}\xspace}
\newcommand{\datasetFinetuningTest}{D\textsubscript{FT-test}\xspace}

\newcommand{\fieldOs}{\textit{OS}\xspace}
\newcommand{\fieldPkgMan}{\textit{Package Manager}\xspace}
\newcommand{\fieldReq}{\textit{Dependencies}\xspace}
\newcommand{\fieldUsesEnv}{\textit{Usage of ENV}\xspace}
\newcommand{\fieldUsesArg}{\textit{Usage of ARG}\xspace}
\newcommand{\fieldUsesLabel}{\textit{Usage of LABEL}\xspace}
\newcommand{\fieldUsesExpose}{\textit{Usage of EXPOSE}\xspace}
\newcommand{\fieldUsesCmd}{\textit{Usage of CMD}\xspace}
\newcommand{\fieldUsesEntrypoint}{\textit{Usage of ENTRYPOINT}\xspace}
\newcommand{\fieldDownExtPkgs}{\textit{Download of External Dependencies}\xspace}

\newcommand{\tfivemodel}[1]{T5$_{\mathit{#1}}$\xspace}
\newcommand{\approach}{\tfivemodel{}}
\newcommand{\baseElastic}{IR$_{\mathit{ES}}$\xspace}
\newcommand{\baseSentT}{IR$_{\mathit{ST}}$\xspace}

\newcommand{\datasetInstancesAll}{670,982\xspace}

\def\BibTeX{{\rm B\kern-.05em{\sc i\kern-.025em b}\kern-.08em
    T\kern-.1667em\lower.7ex\hbox{E}\kern-.125emX}}
\begin{document}

\title{Automatically Generating Dockerfiles \\ via Deep Learning: Challenges and Promises}



\author{
    \IEEEauthorblockN{Giovanni Rosa\IEEEauthorrefmark{1}, Antonio Mastropaolo\IEEEauthorrefmark{2}, Simone Scalabrino\IEEEauthorrefmark{1}, Gabriele Bavota\IEEEauthorrefmark{2}, Rocco Oliveto\IEEEauthorrefmark{1}}
    \IEEEauthorblockA{\IEEEauthorrefmark{1}\textit{STAKE Lab} - University of Molise, Pesche, Italy
    \\\{name.surname\}@unimol.it}
    \IEEEauthorblockA{\IEEEauthorrefmark{2}\textit{Software Institute} - Università della Svizzera Italiana (USI), Switzerland
    \\\{name.surname\}@usi.ch}
}

\maketitle

\begin{abstract}
Containerization allows developers to define the execution environment in which their software needs to be installed. Docker is the leading platform in this field, and developers that use it are required to write a Dockerfile for their software. 
Writing Dockerfiles is far from trivial, especially when the system has unusual requirements for its execution environment. Despite several tools exist to support developers in writing Dockerfiles, none of them is able to generate entire Dockerfiles from scratch given a high-level specification of the requirements of the execution environment.
In this paper, we present a study in which we aim at understanding to what extent Deep Learning (DL), which has been proven successful for other coding tasks, can be used for this specific coding task. 
We preliminarily defined a structured natural language specification for Dockerfile requirements and a methodology that we use to automatically infer the requirements from the largest dataset of Dockerfiles currently available. We used the obtained dataset, with \datasetInstancesAll instances, to train and test a Text-to-Text Transfer Transformer (\approach) model, following the current state-of-the-art procedure for coding tasks, to automatically generate Dockerfiles from the structured specifications. 
The results of our evaluation show that \approach performs similarly to the more trivial IR-based baselines we considered. 
We also report the open challenges associated with the application of deep learning in the context of  Docker file generation.
\end{abstract}


\section{Introduction}
Software companies are more and more often starting adopting the DevOps methodology for developing their products. DevOps strongly relies on technologies for automating the build and the deployment of the systems (CI/CD), and it results in shorter release cycles \cite{humble2010continuous}.
In this context, containerization technologies are fundamental to allow developers take control over the execution environment of their products. Such tools allow developers to reduce the risks of issues arising from possible differences between the development/testing environment and the production environment. Docker is the leading containerization technology, becoming the ``Most Loved" and ``Most Wanted" platform, according to the 2022 Stack Overflow survey \cite{webstacksurvey}.
When using Docker for a specific software product, developers are required to write a Dockerfile, which contains a sequence of instructions that, when executed, allow to build an image (Docker image) which can be run in one or more lightweight virtual machines (Docker containers).

Writing Dockerfiles is not trivial. First, system administration skills are required, which developers do not always have. 

Second, while basic Dockerfiles templates can be used as a starting point, they need to be adapted to the specific requirements of the software system at hand. To clarify this, let us consider the scenario in which a developer needs an environment that contains both Apache Tomcat as a web server and FFMpeg for processing videos, with the support of the x265 codec. 
While templates exist for the former, they do not contain hints about how to provide also the latter. \figref{fig:intro:example} shows an example of possible solution: Such a Dockerfile starts from a pre-defined image which contains Tomcat and it installs FFMpeg with the support of x265 on top of it. It can be noticed that determining the commands required to achieve the latter goal requires a moderate effort and it is prone to errors. This problem has been observed in the literature. For example, in the survey conducted by Reis \etal \cite{reis2021developing}, the authors observed that developers (especially the less experienced ones) perceive the creation of Dockerfiles as a time consuming activity.
Moreover, using Dockerfiles from tutorials and blob posts as a support for their creation can lead to broken Docker images \cite{web:dockerizinghard}.

\begin{figure}
\begin{lstlisting}
FROM tomcat:7.0.75-jre8

RUN echo deb http://archive.ubuntu.com/ubuntu precise universe multiverse >> /etc/apt/sources.list; apt-get update && \
    apt-get -y --fix-missing install autoconf automake build-essential \
    git mercurial cmake libass-dev libgpac-dev libtheora-dev libtool \
    libvdpau-dev libvorbis-dev pkg-config texi2html zlib1g-dev \
    libmp3lame-dev wget yasm && \
    apt-get clean
    
WORKDIR /usr/local/src
# Install x265
RUN hg clone https://bitbucket.org/multicoreware/x265 && \
    cd /usr/local/src/x265/build/linux && \
    cmake -DCMAKE_INSTALL_PREFIX:PATH=/usr ../../source && \
    make -j 8 && \
    make install

WORKDIR /usr/local/src
# Install ffmpeg.
RUN git clone --depth 1 && \
    cd ffmpeg && \
    git://source.ffmpeg.org/ffmpeg && ./configure \
    --extra-libs="-ldl" --enable-gpl --enable-libass \
    --enable-libvorbis --enable-libx265 --enable-nonfree && \
    make -j 8 && \
    make install

WORKDIR /
\end{lstlisting}
\caption{Example of Dockerfile for Tomcat and FFMpeg.}
\label{fig:intro:example}
\vspace{-0.5cm}
\end{figure}

Some tools and approaches have been introduced to help developers to write Dockerfiles.
Some of them \cite{horton2019dockerizeme,nust2019containerit,web:starter,web:boxing,web:dockerfilegenerator} take as input the context and suggest whole Dockerfiles based on it. 
Such tools are handy since they allow developers to quickly have a starting point. They also require limited effort from the developers.\eject

However, they are not able to recognize the need for specific libraries (like FFMpeg, in the previous example). 
Other tools provide a more in-depth support, but they are limited to specific programming languages (\eg DockerizeMe for Python). 
The approach recently introduced by Ye \etal \cite{ye2021dockergen} can recommend packages that need to be installed given an initial set of dependencies, but it only supports developers in the definition of the dependencies to install. The developer still needs to  write complex instructions, like the ones required to build the FFMpeg and x265 libraries in \figref{fig:intro:example}.

Finally, there are code completion tools that support developers while writing Dockerfiles. A recent example is GitHub Copilot \cite{web:copilot}, which relies on Deep Learning (DL) to achieve this goal. Such tools, however, require that developers manually write part of the Dockerfile, so that they can complete it.
None of such tools and approaches is able to generate complete Dockerfiles from a high-level description of what the developer wants in the Dockerfile (\ie software requirements), minimizing the effort required, in terms of writing and knowledge required, to create Dockerfiles.
Previous research \cite{tufano2019empirical, mastropaolo2021studying, mastropaolo2022tse, watson2022systematic} show that DL is a viable solution for code generation-related tasks. However, to the best of our knowledge, no previous work tested to what extent DL can be used to generate complete Dockerfiles.

In this paper, we aim to fill this gap.
We first define the format of a structured high-level specification to define, via natural language, the requirements for the definition of Dockerfiles. Then, we define a methodology for automatically inferring such a \nlRecipeName (\nlRecipe) from existing Dockerfiles, so that we can build a dataset large enough to train and test a DL model. To this aim, we rely on the largest collection of Dockerfiles available in the literature \cite{eng2021revisiting}, containing 9.4M Dockerfile snapshots extracted from \textbf{all} the open-source projects hosted on GitHub. We run our specification-inference tool on them and, after a filtering procedure where duplicates and invalid Dockerfiles are removed, we end up with a set of \datasetInstancesAll unique pairs $\langle$\nlRecipe, Dockerfile$\rangle$. We use this dataset to train and test a state-of-the-art DL model, the Text-to-Text Transfer Transformer (T5) \cite{raffel2020exploring}, which has been proven effective when supporting several coding tasks \cite{mastropaolo2021studying, mastropaolo2022tse},
following the same pipeline defined in the literature.
We compare the DL-based approach with two Information Retrieval (IR)-based approaches (\ie less complex and less-resource-requiring alternatives), and we check to what extent, given a \nlRecipe, the output Dockerfiles of the three techniques: (i) meet the input requirements, (ii) are similar to the target Dockerfile, and (iii) allow to build a Docker image similar to the target one.

We obtain mixed results: While \approach achieves similar results to the best baseline in terms of adherence to the requirements, it generates Dockerfiles less similar to the target Dockerfile. On the other hand, we found that the build of the Dockerfiles generated with \approach succeeds more often and that a higher percentage of intermediate layers produced during the build match the ones obtained with the target Dockerfiles. Interestingly, we also found that \approach truncates the Dockerfiles. 
\eject

This might be due to two main issues. 
First, \textbf{a larger training dataset might be needed for this task}: Despite we consider the largest collection of Dockerfiles in the literature \cite{eng2021revisiting}, our results suggest that the T5 learning could benefit from more training data, considering the number of used instances to previous studies working on source code~\cite{mastropaolo2021empirical}. 
Second, \textbf{a different \approach training stop criterion needs to be defined}: The stop criterion we adopt, which is the one currently used for coding tasks \cite{mastropaolo2022tse}, is based on the convergence in terms of BLEU-4 score. However, considering our results, it seems to be ineffective in the evaluated context.


\newlength\MAX  \setlength\MAX{15mm}
\newcommand*\Chart[1]{\rlap{\textcolor{black!20}{\rule{\MAX}{2ex}}}\rule{#1\MAX}{2ex}}
\begin{table*}[htp]
 \centering
 \caption{Format of \nlRecipes, with the type of each field, a description, and the percentage of survey participants who indicated the field as important.}
 \label{tab:nlrecipes}
 \resizebox{\linewidth}{!}{
 \begin{tabular}{l l p{11cm} l}
  \toprule
  \textbf{Field}       & \textbf{Type}  & \textbf{Description}                                                                                                                              & \textbf{\% Positive Answers} \\
  \midrule                                                                                                                                                                                   
  \fieldOs	           & String         & Operating system to be used.                                                                                                                      & 75\%~\Chart{0.75}   \\
  \fieldPkgMan         & String         & Linux package manager to be used (\eg \texttt{apt} or \texttt{apk}).                                                                              & 58\%~\Chart{0.58}   \\
  \fieldReq		       & String[]       & Dependencies (\eg packages) that must be provided.                                                                                                & 83\%~\Chart{0.83}   \\
  \fieldDownExtPkgs    & Boolean        & Dependencies can be installed as external resources (\ie not through the package manager).                                                        & 58\%~\Chart{0.58}   \\
  \fieldUsesEnv        & Boolean        & Environment variables must be supported for customizing the container.                                                                            & 83\%~\Chart{0.83}   \\
  \fieldUsesArg        & Boolean        & Arguments must be supported for customizing the image build process.                                                                              & 75\%~\Chart{0.75}   \\
  \fieldUsesLabel      & Boolean        & Labels must be used for documenting the Dockerfile.                                                                                               & 50\%~\Chart{0.50}   \\
  \fieldUsesExpose     & Boolean        & Network ports used must be documented.                                                                                                            & 67\%~\Chart{0.67}   \\
  \fieldUsesCmd        & Boolean        & \multirow{2}{11cm}{The command to execute when starting the container must be specified through the \texttt{CMD} or \texttt{ENTRYPOINT} command.} & 75\%~\Chart{0.75}   \\
  \fieldUsesEntrypoint & Boolean        &                                                                                                                                                   & 75\%~\Chart{0.75}   \\
  \bottomrule
 \end{tabular}
 }
\end{table*}

\begin{figure*}[htp]
    \centering
    \includegraphics[width=\linewidth]{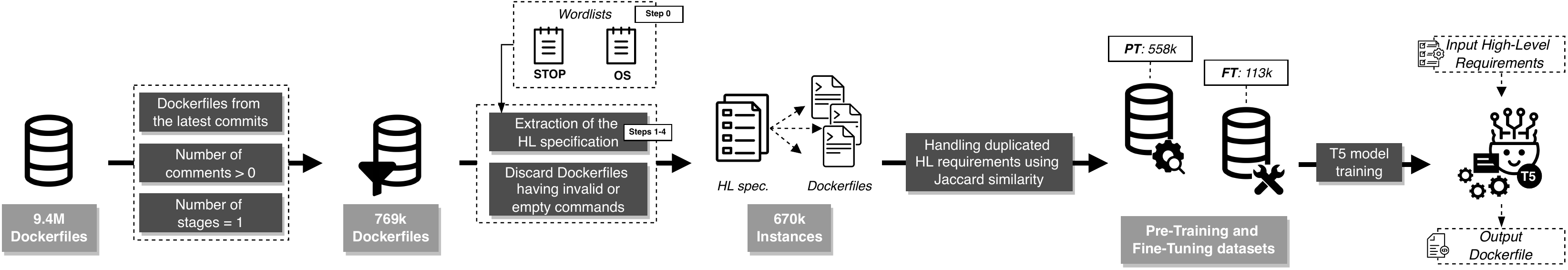}
    \caption{Steps performed to train \approach for generating Dockerfiles from specifications.}
	\label{fig:workflow}
	\vspace{-0.3cm}
\end{figure*}


\section{Background \& Related Work}
Writing a Dockerfile is the first step to containerize an application using Docker.
A Dockerfile specifies the dependencies and execution environment to build and execute the target application using domain-specific instructions \cite{web:dockerreference}.
The first instruction is the \texttt{FROM}, which defines on which existing image the new image builds upon, \ie the \textit{base image}.
Then, we have instructions that perform actions during the build, such as \texttt{RUN}, to execute scripts, \texttt{WORKDIR}, to change the working directory, and \texttt{COPY/ADD}, to copy files, folders and compressed archives.
Other instructions are for metadata and documentation (\texttt{LABEL}, \texttt{EXPOSE}), build arguments (\texttt{ARG}), environment variables (\texttt{ENV}), and commands that the container must execute when the image is run (\texttt{CMD}, \texttt{ENTRYPOINT}).
Each Docker image built from a Dockerfile is uniquely identified by a \textit{digest}, a hash value computed upon the files that compose the image.
While building a Dockerfile, Docker creates a \textit{layer} for each instruction, which constitutes a temporary image that allows to cache and speed-up the next build by avoiding the build of previously-built layers, when possible.
The layers are also identified by a unique \textit{digest}.

Several tools and approaches are available to help developers in writing Dockerfiles. Due to space constraints, we only focus on such approaches, which are the most related to our work.
\textit{GitHub Copilot} \cite{web:copilot} has been recently introduced as a general-purpose code completion tool, and it works also with Dockerfiles. \textit{Humpback} \cite{hanayama2020humpback} achieves a similar goal, but it is specifically designed for Dockerfiles. The tool by Zhang \etal \cite{zhang2022recommending} helps developers choosing the right \textit{base image}.
Other approaches provide more advanced support.
Horton \etal \cite{horton2019dockerizeme} proposed \textsc{DockerizeMe}, an approach for the automatic inference of environment dependencies starting from Python source code, without requiring inputs from developers.
However, it only targets the execution dependencies for Python code.
Ye \etal \cite{ye2021dockergen}, proposed \textsc{DockerGen}, an approach that uses knowledge graphs, built upon 220k Dockerfiles, for generating Dockerfiles for a specific software application.
Starting from a target software, \textsc{DockerGen} infers all the dependencies required for the execution environment including the selection of a suitable base image.
However, the support provided by \textsc{DockerGen} is limited to the recommendation of dependencies, while we aim at generating complete Dockerfiles (including, for example, the \texttt{RUN} instructions required for building an external package, as the ones in \figref{fig:intro:example}).

Other tools take as input the context (\ie the root folder of the project) and suggest whole Dockerfiles based on it. 
An example is the tool \textit{starter} \cite{web:starter}, that generates a Dockerfile and a \textit{docker-compose.yml} file from arbitrary source code.
Some other tools provide a more in-depth support for specific programming languages, based on the project context. 
For example, there are tools specific for R \cite{nust2019containerit}, Node.js \cite{web:dockerfilegenerator}, Ruby \cite{web:boxing}, and PHP \cite{web:phpdockerio}. Others support multiple languages, such as \textit{generator-docker} \cite{web:generatordocker}.
To the best of our knowledge, no previous work introduced an approach that is able to generate complete Dockerfiles given high-level requirements.

\newcommand{\stepZeroFROMs}{10,960,563\xspace}
\newcommand{\stepZeroUniqueWords}{46,070\xspace}
\newcommand{\stepZeroSelectedWords}{2,120\xspace}
\newcommand{\stepZeroOS}{74\xspace}
\newcommand{\stepZeroStopwords}{920\xspace}

\newcommand{\wordlistStop}{WL\textsubscript{stop}\xspace}
\newcommand{\wordlistOs}{WL\textsubscript{OS}\xspace}
\newcommand{\wordlistExclude}{WL\textsubscript{to-exclude}\xspace}

\newcommand{\dockerfileParsed}{769,385\xspace}

\newcommand{\datasetPretraining}{D\textsubscript{PT}\xspace}
\newcommand{\datasetFinetuning}{D\textsubscript{FT}\xspace}

\section{Deep Learning for Generating Dockerfiles}
\label{sec:approach}

We define a procedure to train a DL model for the generation of Dockerfiles from \nlRecipeNames. 

In \figref{fig:workflow} we report the workflow we used to train a DL-based model (\approach) for the generation of Dockerfiles.
We first define a structured \nlRecipeName (\nlRecipe) for Dockerfiles. Then, we extract \nlRecipes from existing Dockerfiles through an automated approach.
Finally, we use such a dataset to train the \approach model for the generation of Dockerfiles (\nlRecipe $\rightarrow$ Dockerfile).
In the following, we describe in detail the steps to construct our model.

\subsection{Dockerfile High-Level Specification}

Natural language can be used as an effective tool for reporting the requirements of the source code. When it comes to Dockerfiles, however, the high-level requirements that can be expressed are much more limited. For the source code, a developer might want to specify, for example, constraints on the input parameters and conditions that lead to errors. On the other hand, for Dockerfiles, it boils down to a matter of what the developer wants installed in the container, plus a few more characteristics.

Thus, to standardize the format of a Dockerfile requirements specification, written in natural language, the idea is to define a set of key-value requirements.
In a real-world application, it could be seen as a structured form where, for each field, the developer specifies the values to meet the requirements.
Based on the commands available in Dockerfiles and on how Dockerfiles are generally structured (based on our experience), we distilled a structured format for \nlRecipeNames (\nlRecipes), which we report in \tabref{tab:nlrecipes}.

We did not include in the specification requirements related to particular Dockerfile commands (\ie \texttt{ADD}, \texttt{COPY}, \texttt{HEALTHCHECK}, \texttt{MAINTAINER}, \texttt{ONBUILD}, \texttt{SHELL}, \texttt{STOPSIGNAL}, \texttt{USER}, \texttt{VOLUME}, \texttt{WORKDIR}) for three reasons: First, they are related more to low-level details (\eg how the user of the Dockerfile should be set up); Second, in general, developers do not frequently use all of them \cite{9cito2017empirical,eng2021revisiting}; Third, some of them are deprecated (\ie \texttt{MAINTAINER}). Ideally, our approach automatically generates such low-level instructions when needed, \ie based on the instruction needed by the application that must be containerized.
Also, developers might still easily tune them up on the generated Dockerfile, if they want to.

To validate the specification structure, we ran a survey in which we asked 12 professional software developers with at least 2 years of experience with Docker whether they would want to specify each of the fields we hypothesized to be relevant in a \nlRecipe (binary yes/no question). The respondents come from personal invitation and social media, with a $\sim$40\% response rate. We report the results in \tabref{tab:nlrecipes}. All the fields are important to at least 50\% of the participants. The least relevant field, according to the participants, is \fieldUsesLabel (50\% positive answers), while the most important ones are \fieldReq and \fieldUsesEnv (83\% positive answers).

\subsection{Inferring Requirements from Existing Dockerfiles}
\label{sec:nlparser}
We need a large amount of associations \nlRecipe $\rightarrow$ Dockerfile, to train a DL-based Dockerfile generation model. While Dockerfiles are largely available, this does not hold for the requirements behind them. The latter could be manually inferred, but such a process would be infeasible for a large-scale dataset. Thus, we defined an automated procedure for inferring the \nlRecipe behind an existing Dockerfile.
In summary, the process works as follows. Given a list of operating systems (\wordlistOs) and stop words (\wordlistStop), we extract the operating system (step 1), the software dependencies (step 2), and, finally, the other fields required for \nlRecipe (step 3).
In the following, we describe in detail each step of our methodology, and the procedure we used to define the two previously-mentioned lists of words.

\subsubsection{Step 1: Inferring the OS} 
To infer the OS required by the developer, we only focus on the \texttt{FROM} instruction, which comes in the format \texttt{FROM <name>[:<tag>]}.
Since \texttt{<name>} and \texttt{<tag>} are usually composed by one or more words, we extract them by splitting their content of image name and tag by the typically employed as separators, \ie \texttt{-} and \texttt{\_}. 
We first check if any word in the \wordlistOs is present in the words extracted from the tag, and then in the name. If a OS-related keyword is present, we set \fieldOs with such a keyword, while we use the special keyword ``any'' otherwise, indicating that the developer did not have any requirement in terms of operating system. If more than a OS-related keyword is present, we consider the first match found. For example, for the instruction \texttt{FROM tomcat:9.0.20-jre8-alpine}, we extract the keywords (in the order): ``9.0.20'', ``jre8'', ``alpine'' (from the tag), and ``tomcat'' (from the image name). ``alpine'' is the first (and only, in this example) OS-related word. Therefore, we set \fieldOs to ``alpine''.
If the first OS-related keyword is found in the image name, we also add to it all the keywords containing only numbers from the tag name, which most likely refer to the version number. For example, for the instruction \texttt{FROM debian:10-slim}, we set \fieldOs to ``debian10''.

\subsubsection{Step 2: Inferring Software Dependencies}
Software dependencies might be present in several Dockerfile instructions. The first instruction which might contain dependencies is the \texttt{FROM} instruction. To detect the (possible) dependencies explicitly expressed in such a field, we extract all the words from the image name (with the same procedure described in step 1), we remove the words in \wordlistOs (OSes) and \wordlistStop (stop words), and we exclude the non-alphabetic words. 

All the remaining words are added to the list of software dependencies (\fieldReq). In the previous example, \texttt{FROM tomcat:9.0.20-jre8-alpine}), the dependency extracted (and only word used as image name) is ``tomcat''.
At this point, we need to extract all the other dependencies installed with package managers or other procedures (\eg downloaded and installed). This task is far from trivial. Simply considering the packages installed through package managers (\eg \texttt{apt install}) is not an option: Most of the packages installed do not correspond to high-level requirements, but rather to low-level details about support libraries. 

For example, in \figref{fig:intro:example}, the package \texttt{build-essential} is not a dependency of the software system, but rather a package incidentally needed (in this case, for building two actual dependencies, \ie \texttt{x265} and \texttt{ffmpeg}). We want our DL model to automatically infer the need of such packages.

However, it is known that developers tend to give an explanation comment of what each Dockerfile instruction does. Thus, we can use those comments to extract only high-level requirements, which are reported by the developers themselves.
To achieve this, we use the following heuristic. We first select all the comment lines that contain the word ``install''. We tokenize each comment with the \textit{spacy} Python library \cite{web:spacy}, and we detect the words that depend on such a keyword. We discard from the obtained list all the words in \wordlistStop, and we select the remaining ones as candidate dependencies for the specific comment line.
For each comment line $c_i$ with its candidate list of dependencies $d_{c_i}$, we process each \texttt{RUN} instruction between $c$ and the next comment line ($c_{i+1}$) or blank line. 
We use \textit{bashlex} \cite{web:bashlex} for parsing the bash script in each RUN instruction, and we split it in statements. Finally, we consider only the statements in which the most common package managers (\ie \texttt{apt}, \texttt{yum}, \texttt{apk}, \texttt{pip}, and \texttt{npm}) and commands for downloading files (\ie \texttt{curl} and \texttt{wget}) are present, and we extract their arguments. If there any argument (packages) matches a candidate requirement (from the comment), we add the candidate requirement in \fieldReq. Note that we do not simply consider the words depending on the ``install'' word from the comments because some words might not be software dependencies (\eg in ``install only ruby,'' the word ``only'' should be ignored).
Finally, we exclude duplicates, and we set the \fieldReq with the dependencies extracted both from the \texttt{FROM} instruction and from the \texttt{RUN} instructions.

\subsubsection{Step 3: Parsing Additional Fields}
All the remaining fields are straightforward to be set. For the fields \fieldUsesArg, \fieldUsesCmd, \fieldUsesEntrypoint, \fieldUsesEnv, \fieldUsesExpose, and \fieldUsesLabel, we simply check if at least one of the respective instructions is present in the Dockerfile.
The field \fieldPkgMan is defined by checking if any of the most commonly used package managers for Linux distributions (\ie \texttt{apt}, \texttt{apk}, and \texttt{yum}) are used in the \texttt{RUN} instructions. If this is the case, we set \fieldPkgMan to such a package manager (\eg \fieldPkgMan = \texttt{yum}). We also check for the coherence between the package manager and the OS (Linux distribution) detected in step 1: For example, if the OS is \texttt{ubuntu}, the package manager can not be \texttt{yum}. If this happens, or if no specific package manager is detected (\eg no package is installed), we set \fieldPkgMan to \texttt{any}.
Finally, for the field \fieldDownExtPkgs, we check if any \texttt{RUN} instruction contains one of the following: (i) a link in the context of a download-related instruction (\eg \texttt{wget} or \texttt{git clone}); (ii) the installation of a Python or JavaScript external library; (iii) the installation of an external package through the package managers (\eg \texttt{dpkg} for Debian/Ubuntu). If this happens, we set \fieldDownExtPkgs to \texttt{true} (\texttt{false} otherwise).

\subsubsection{Defining OSes and Stop Words}
We use a systematic procedure for defining the two lists required by the \nlRecipe parser, \ie \wordlistOs and \wordlistStop.
We extracted all the \texttt{FROM} instructions contained in the Dockerfiles from the collection by Eng \etal \cite{eng2021revisiting}, that we later use to build our dataset. In total, we obtained \stepZeroFROMs instructions.
Then, we extract keywords from the image name and tag, like we do in step 1.
Next, we discard the words that (i) contain only non-alphabetic characters (\eg version numbers), or (ii) have less than 3 characters (most likely stop words).
We obtain a list of \stepZeroUniqueWords unique words, along with the respective count of occurrences. We filter out all the words with less than 150 occurrences, thus obtaining \stepZeroSelectedWords words, covering 97\% of the total occurrences of all the words extracted.
We manually analyzed and labeled each of them as ``OS'', ``stop word'', or ``dependency''. For example, ``centos'' is marked as ``OS'' keyword, ``baseimage'' as ``stop word'', while ``python'' as ``dependency''. 
In the end, we obtained \stepZeroStopwords stop words for \wordlistStop and \stepZeroOS words referring to OSs for \wordlistOs.

\subsection{Defining a Dataset of \nlRecipes and Dockerfiles}
\label{sec:nldataset}
We proceed with using our parser to build the dataset of \nlRecipes and target Dockerfiles.
We use the dataset built by Eng \etal \cite{eng2021revisiting}, which is the largest (9.4M) and the latest dataset of Dockerfiles currently available in the literature. That dataset comes from the S version of World of Code (WoC) \cite{ma2019world}, covering a period of time ranging between 2013 and 2020. To the best of our knowledge, this is the most recent and large collection of Dockerfiles available in the literature. The dataset provides all the versions (snapshots) of all the Dockerfiles extracted from GitHub projects, along with the related commit ID. For building our dataset, we only select the latest commit for each repository.
As a result, we obtain 3,010,141 Dockerfiles.

We filter out all the Dockerfiles that have no comments (required in step 2 of the parser), and also those that have more than one stage, as they are currently not supported by our parser.
Next, we drop all the duplicated Dockerfiles (according to their \textit{sha1} hash) and all the Dockerfiles that contain in the \texttt{FROM} instruction at least a keyword that we did not manually evaluate when defining the two lists \wordlistOs and \wordlistStop, to ensure that we do not unintentionally include stop words as dependencies.
After applying these filters, we obtain a total of \dockerfileParsed Dockerfiles, on which we run our approach to extract the \nlRecipes.

At this stage, we further exclude all the Dockerfiles for which our parser detected syntax errors in bash instructions and empty Dockerfile instructions (\eg \texttt{COPY} without arguments).
As a result, we obtain a total of \datasetInstancesAll pairs of Dockerfile, associated with the respective \nlRecipes (in total, we found 121,030 unique \nlRecipes). At this stage, each \nlRecipe is associated with one or more Dockerfiles since different Dockerfiles might result from the same requirements. We need, however, to select a single representative Dockerfile for each \nlRecipe that we can use for the training and the test of \approach. To do this, we first select all \nlRecipe associated with two or more Dockerfiles. We found 41,820 of them. 

Given a set of Dockerfiles associated with the same \nlRecipe, we select the one containing the highest number of ``typical'' instructions for that cluster, to obtain the most typical Dockerfile for the given \nlRecipe. We tokenize each instruction of all the Dockerfiles. Then, given two Dockerfiles $A$ and $B$, we compute the Jaccard similarity between each pair of instructions of the same kind (\eg \texttt{COPY} instructions are compared only with \texttt{COPY} instructions), with the formula $J(W_{A_i}, W_{B_j}) = \frac{|W_{A_i} \cap W_{B_j}|}{|W_{A_i} \cup W_{B_j}|}$, where $W_{A_i}$ and $W_{B_j}$ are the words from instructions $i$ and $j$ of $A$ and $B$, respectively. We choose as representative the Dockerfile with the highest mean similarity over all the instructions.

We further process the Dockerfiles to prepare them for training: First, given all the package installation instructions (\eg \texttt{apt install}), we sort the packages in lexicographic order, to avoid that different package orders confuse the DL approach. Second, we remove all the lines that contain only comments, to avoid that the model makes extra efforts in solving a sub-problem which is not in the scope of this paper.

From this, we extract two sub-datasets: \datasetPretraining for the pre-training, and \datasetFinetuning for the fine-tuning of the T5 model.
As for the former, we use all the Dockerfiles discarded while choosing the representative Dockerfiles for each \nlRecipe and all the Dockerfiles that are associated with a \nlRecipe for which the field \fieldReq is empty (no software dependency needs to be installed). It is most likely that such Dockerfiles do have dependencies, but we were not able to find them because of the lack of comments in the format we expected (\eg the word ``install'' has not been used in comments).
All the remaining pairs $\langle$\nlRecipe, Dockerfile$\rangle$ are placed in \datasetFinetuning.
In the end, we obtain 557,540 instances for \datasetPretraining and 113,442 instances for \datasetFinetuning. As a requirement for the training of T5, we can only use Dockerfiles having no more than 1024 token, obtaining a total of 113,131 instances.
Then, we divide the \datasetFinetuning in training- (\datasetFinetuningTrain), evaluation- (\datasetFinetuningEval), and test- (\datasetFinetuningTest) sets, by performing a typical 80\%-10\%-10\% spliting \cite{mastropaolo2021studying, mastropaolo2022tse}, obtaining 90,504, 11,313, and 11,314 instances, respectively. We use \datasetFinetuningTrain for fine-tuning \approach, \datasetFinetuningEval for the hyper-parameter tuning, and \datasetFinetuningTest for our experiment (see \secref{sec:design}).

\subsection{Training T5 for Generating Dockerfiles}
\label{sec:model}
Raffel \etal \cite{raffel2020exploring} introduced \approach to support multitask learning in Natural Language Processing. Such a model re-frames NLP tasks in a unified text-to-text format in which the input and output of all tasks are always text strings. A T5 model is trained in two phases: (i) \textit{pre-training}, in which the model is trained with a self-supervised objective that allows defining a shared knowledge-base useful for a large class of tasks, and \textit{fine-tuning}, which specializes the model on a downstream task (\eg language translation).
T5 already showed its effectiveness in code-related tasks \cite{mastropaolo2021studying, mastropaolo2021empirical, mastropaolo2022lance, tufano2022using, li2022auger, zhang2022coditt5, ciniselli2021empirical}. However, its application to the generation of Dockerfiles is novel and still unexplored.
As done in previous work \cite{mastropaolo2021studying, mastropaolo2021empirical}, we use the smallest T5 version available (T5 small), which is composed of 60M parameters.
\eject

Given a prediction provided by the model, the output token streams can be generated using several decoding strategies. We use \textit{greedy decoding} when generating an output sequence. In detail, such a strategy selects, at each time step $t$, the symbol having the highest probability of appearing in a specific position. 
We describe below both the pre-training and the fine-tuning procedure we applied for this task.

\subsubsection{Pre-Training Procedure}
\label{sub:pretraining}
The ``general knowledge'' \cite{raffel2020exploring} that we want to provide our model with is, in our case, a mixture of technical natural language (English) and technical language (Dockerfiles). We experiment with three pre-training variations: \tfivemodel{NL}, which only relies on natural language, \tfivemodel{DF}, which only relies on Dockerfiles, and \tfivemodel{NL+DL}, which relies on both. We test all such three variants and we pick the best after performing hyper-parameter tuning.

As for the first variant, \tfivemodel{NL}, we use the pre-trained checkpoint \cite{t5-checkpoint} released by Raffel \etal \cite{raffel2020exploring}. We do not perform any further pre-training for such a model. 

Instead, we leverage the knowledge that has been already gained when pre-training the T5 model on the English text (C4 corpus \cite{raffel2020exploring}) for 1M steps.

As for the second variant, \tfivemodel{DL}, we adopt a classic \emph{masked language model} task, \ie we randomly mask 15\% of the tokens in a training instance, asking the model to predict them. We pre-train such a model on \datasetPretraining. 
Finally, as for the third variant, \tfivemodel{NL+DF}, we start from the \tfivemodel{NL} model and we further pre-train it for 500k steps on \datasetPretraining, using the same procedure used for pre-training \tfivemodel{DF}.
Finally, we created a new \emph{SentencePiece} model \cite{kudo2018sentencepiece} for tokenizing natural language text. We trained it on \datasetPretraining.
For both the models for which we performed additional pre-training steps (\tfivemodel{DF} and \tfivemodel{NL+DF}), we used a 2x2 TPU topology (8 cores) from Google Colab with a batch size of 16 and a sequence length of 512 tokens for the input and 1,250 for the output. As a learning rate, we use the Inverse Square Root with the default configuration \cite{raffel2020exploring}. For the pre-training phase, we use the default parameters defined for the T5 model \cite{raffel2020exploring}.  

\subsubsection{Hyper-parameter Tuning} 
We test four learning rate strategies, \ie constant learning rate (C-LR), slanted triangular learning rate (ST-LR), inverse square learning rate (ISQ-LR), and polynomial learning rate (PD-LR). We report in \tabref{tab:learning-rates} the parameters we use for each of them. 

Given the three pre-trained models, \tfivemodel{NL}, \tfivemodel{DF}, and \tfivemodel{NL+DF}, we fine-tune them on \datasetFinetuningEval (100k steps, batch size of 32, input sequence length of 512 tokens, output sequence length of 1024 tokens), leading to 12 different models (3 models $\times$ 4 strategies). Then, to assess their performance, we compute the BLEU-4 \cite{papineni2002bleu} metric between the generated Dockerfiles and the target ones. Such a metric has been used in previous work for other coding tasks \cite{liu2018neural,leclair2019neural,mastropaolo2021studying, mastropaolo2022tse} and it ranges between 0 (completely different) and 1 (identical).
We report in \tabref{tab:hp-results} the results achieved by the 12 models in terms of BLEU-4.
The best results are achieved with \tfivemodel{DF} with the ISQ-LR strategy and \tfivemodel{NL+DF} with the ST-RL strategy (17.20\% BLEU-4 for both). 

In the end, we select the latter since \tfivemodel{NL+DF} achieves better results also for the other strategies. 
\begin{table}
	\centering
	\caption{Configurations for the experimented learning rates}
	\label{tab:learning-rates}
	\begin{tabular}{ll|ll}   
	  \toprule
	  \textbf{Strategy}    & \textbf{Parameters}                   & \textbf{Strategy}    & \textbf{Parameters}     \\
	  \midrule                                                     
	  C-LR            & \textit{LR = 0.001}                        & ISQ-LR  & \textit{LR$_s$ = 0.01}   \\
	  ST-LR            & \textit{LR$_s$ = 0.001}                   &                      & \textit{W = 10,000}      \\
						  & \textit{LR$_{max}$ = 0.01}             & PD-LR     & \textit{LR$_s$ = 0.01}   \\
						  & \textit{Ratio = 32}                    &                      & \textit{LR$_e$ = 0.001}  \\
						  & \textit{Cut = 0.1}                     &                      & \textit{Pow = 0.5}     \\  
	  \bottomrule
	\end{tabular}
\end{table}

\subsubsection{Fine-tuning}
We fine-tune the best pre-trained model (\tfivemodel{NL+DF}) with the best learning rate strategy (ST-LR) on \datasetFinetuningTrain. We use early stopping to avoid overfitting \cite{yao2007early, tufano2022using}: We save a checkpoint every 10k steps and compute the BLEU-4 score on the evaluation set every 100k steps. When the 100k steps do not lead to an improvement, we stop the training procedure, and we keep the last model. 

\begin{table}
	\centering
	\caption{T5 hyper-parameter tuning results (BLEU-4).}
	\begin{tabular}{lrrrr}
		\toprule
		\textbf{Experiment}                  																		& \textbf{C-LR}              & \textbf{ST-LR}      & \textbf{ISQ-LR}        & \textbf{PD-LR} \\
		\midrule
		\tfivemodel{NL}                          &   13.80\%                & 13.70\%    		           & 5.50\%           &  \bf 14.50\%         \\
		\tfivemodel{DF}                        &   16.90\%                & 5.50\%    		           & \bf 17.20\%          &  15.90\%         \\
		\tfivemodel{NL+DF}                         &  16.60\%                & \bf 17.20\%    	&	           16.50\%           &  17.10\%         \\
		\midrule
	\end{tabular}
	\label{tab:hp-results}
	\vspace{-0.3cm}
\end{table}

\newcommand{\trainingSetInstances}{90,504\xspace}
\newcommand{\testSetInstances}{11,314\xspace}

\newcommand{\instancesRqOneDL}{9,977\xspace}
\newcommand{\instancesRqOneElasticSentT}{11,314\xspace}

\newcommand{\instancesRqTwoDL}{9,963\xspace}
\newcommand{\instancesRqTwoElastic}{11,311\xspace}
\newcommand{\instancesRqTwoSentT}{11,129\xspace}

\newcommand{\testSetInstancesRqThree}{4,059\xspace}
\newcommand{\testSetReposRqThree}{3,909\xspace}
\newcommand{\testSetSampleRqThree}{500\xspace}

\section{Empirical Study Design}
\label{sec:design}

The \textit{goal} of our study is to understand to what extent \approach is effective in generating Dockerfiles.
Our study is steered from the following research questions:
\begin{itemize}[itemindent=0.2cm]
  \item[\RQ{1}:] \textit{To what extent is \approach able to generate Dockerfiles meeting the input natural language specification?}    We evaluate the effectiveness of \approach in generating Dockerfiles that meet the requirements.

  \item[\RQ{2}:] \textit{To what extent are the Dockerfiles generated by \approach similar to the original ones written by developers?}    With this second RQ, we aim at understanding if \approach generates Dockerfiles similar to the targets ones.

  \item[\RQ{3}:] \textit{To what extent are the Docker Images built from the Dockerfiles generated by \approach similar to the original ones built form the Dockerfiles written by developers?} 
  Two Docker images can be equal and come from completely different Dockerfiles. Therefore, with this last RQ, we try to understand whether the images built from the generated Dockerfiles are similar to the original ones.
\end{itemize}

\subsection{Baseline Techniques}
We use as baseline techniques two Information Retrieval (IR)-based approaches.
The first one is \baseElastic, and it is based on the state-of-the-practice for implementing IR approaches and search engines, \ie Elasticsearch \cite{web:elastic}. Given a collection of documents ($D$) and a query ($q$), Elasticsearch first assigns a score to each document in $D$ according to $q$ and then it sorts them. The score is computed with Okapi BM25 \cite{web:bm25}. We add into an Elasticsearch instance all the instances in \datasetFinetuningTrain. Specifically, for each instance, we define a document that contains both the \nlRecipe and the associated Dockerfile. Given a new \nlRecipe for which we want to get a candidate Dockerfile, we perform a \textit{boolean query} composed of all the fields of the \nlRecipe in OR clause (\ie \texttt{should}, in Elasticsearch). 
We report an example of the query in our replication package \cite{replication}.

The second baseline we consider is \baseSentT, and it is based on the \textit{SentenceTransformers} framework \cite{reimers2019sentencebert}. Such a framework allows to train embeddings for several data types (including text) so that they can be represented as numeric vectors. First, we use \datasetFinetuningTrain to train the model for computing the embeddings (\textit{bert-base-uncased}). Then, we compute the embeddings for each \nlRecipe as $e(d_{\mathit{Spec}})$ for each \nlRecipe in \datasetFinetuningTrain, and we store their associations with the respective Dockerfiles ($e(d_{\mathit{Spec}}) \rightarrow d_{\mathit{Dockerfile}}$). Given a new \nlRecipe, $t_{\mathit{Spec}}$, we compute its embeddings ($e(t_{\mathit{Spec}})$) and, then, the cosine similarity between $e(t_{\mathit{Spec}})$ and each $e(d_{\mathit{Spec}})$. Finally, we return the $d_{Dockerfile}$ for which the aforementioned similarity is maximum.

\subsection{Context Selection}
The \textit{context} of our study is composed of (i) a set of associations \nlRecipe $\rightarrow$ Dockerfile, (ii) the Dockerfiles generated/retrieved by \approach and the two baseline techniques, and (iii) the source code of the software projects for which the Dockerfiles need to be built, for building the images and thus answering \RQ{3}.

As for the first object, we used the \datasetFinetuningTest dataset.
To obtain the second object, we ran \approach, \baseElastic, and \baseSentT on the \nlRecipes from \datasetFinetuningTest. As a result, we obtained three sets of generated/retrieved Dockerfiles, \ie DF$_{T5}$, DF$_{ES}$, DF$_{ST}$.
Finally, to obtain the third object, for each Dockerfile in \datasetFinetuningTest, we consider the original entry in the dataset by Eng \etal \cite{eng2021revisiting} and we recover the project from which it was extracted and the commit for that specific snapshot.
We cloned all the repositories corresponding to the test instances, discarding the ones for which the source commit or repository was no longer available.
As a result, we obtained a total of \testSetReposRqThree repositories, corresponding to \testSetInstancesRqThree instances of \datasetFinetuningTest. Since building Dockerfiles requires a large amount of time, we did this for a representative sample of \testSetSampleRqThree Dockerfiles from \datasetFinetuningTest (4.28\% margin of error, 95\% confidence level). For each instance in \datasetFinetuningTest, we tried to build the original Dockerfile: If the build failed, we discarded the instance, while, if it succeeded, we kept it, until we collected \testSetSampleRqThree instances. 

\subsection{Experimental Procedure}
To answer \textbf{\RQ{1}}, we compare the \nlRecipe related to the Dockerfiles returned by the approaches we consider with the one given as input.
As for the two baselines, we already have an associated \nlRecipe for each returned Dockerfile (\ie the one from \datasetFinetuningTrain). This, however, is not true for \approach since it generates Dockerfiles from scratch. In this case, we use the same process described in \secref{sec:nlparser} on the generated Dockerfiles for all the fields, except for \fieldReq. We check if requirements in the input \nlRecipe are met in the generated Dockerfile. The reason is that we trained \approach not to generate comments: Since our procedure for extracting \fieldReq strongly relies on comments, we can not directly use it on the generated Dockerfiles. 
We ignore the Dockerfiles generated by \approach for which the parser we defined is not able to infer all the fields in the \nlRecipe (1,337 instances, \ie $\sim$12\% of the total). We consider, instead, all of them for the two baselines, for which this cannot happen since, as mentioned, we do not infer the \nlRecipe.
Finally, we measure the similarity between the target \nlRecipes and the obtained ones.
To achieve this, we assign a \textit{score} for each field of the \nlRecipes, compared to the respective instance in the target \nlRecipe, which is computed by assigning 1 point if the field is equal, and 0 otherwise.
The only exception is the \fieldReq field: Since this is a collection of elements, in this case we compute the score by computing the percentage of elements in the target \nlRecipe that are present also in the obtained one (\ie the recall).
We compute and report the mean score obtained for each field of the \nlRecipes. 


To answer \textbf{\RQ{2}}, we compare the sets DF$_{T5}$, DF$_{ES}$, DF$_{ST}$ with the respective target Dockerfiles. Simply computing textual similarity for Dockerfiles is not sufficient since, in many cases, instructions can be swapped without affecting the final result. Therefore, we rely on the AST representation of the Dockerfiles. To do this, we use the \textit{binnacle} tool by Henkel \etal \cite{19henkel2020learning}. Binnacle extracts ASTs at three different abstraction levels. We use the \textit{phase-2} representation. We do not use the \textit{phase-3} abstraction level since it abstracts the bash commands. For example, both the \texttt{apt-get install} and \texttt{pacman -S} instructions get replaced with a generic \textit{package install}. While such an abstraction is useful, the tool does not support all the possible bash commands, thus causing loss of information, in some cases. 
We report an example of a parsed AST in our replication package \cite{replication}.
Given the ASTs of two Dockerfiles, we compute the \textit{edit distance} between them, \ie the number of modifications needed to transform one into the other, using the Zhang-Shasha algorithm \cite{zhang1989simple}. We normalize the obtained edit distance by dividing it by the sum of the sizes of the two trees we are comparing.
We discard all the instances for which the \textit{binnacle} tool is not able to extract the AST. We obtain \instancesRqTwoDL for \approach, \instancesRqTwoElastic for \baseElastic, and \instancesRqTwoSentT for \baseSentT.
We run the Mann–Whitney U test \cite{wilcoxon1945individual} to compare \approach with the baselines in terms of normalized edit distance. The null hypothesis is that there is no difference between the edit distance obtained using \approach and the one obtained using an IR-based approach. We correct for multiple comparisons using the Benjamini-Hochberg procedure \cite{benjamini1995controlling}.
Finally, we compute the effect size using Cliff's Delta \cite{cliff1993dominance}, which is negligible for all of the performed comparison (\ie \approach compared with the two baselines).
This means that their difference is negligible, even if it is statistically significant.

To answer \textbf{\RQ{3}}, we compare the Docker images built from the resulting Dockerfiles provided by the three techniques and the one built from the target Dockerfile. We consider the GitHub projects we cloned for the sample of \testSetReposRqThree instances and, for each of them, we replace the original Dockerfile with the one generated/retrieved by the three approaches, one at a time, and we try to build it. For each instance, we memorized (i) if the build succeeded, (ii) if the the original and the obtained images are equal, and, if not, (iii) to what extent the latter provides what is also present in the former.
As for the second measurement, we rely on the image digest: We say that two images are equal if their digest are equal. While this is not true by design, we can safely assume that, in our context, the risk of obtaining different images with the same digest is negligible.
As for the third measurement, we compute the digest of each build layer (\ie one for each Dockerfile instruction), and we compute the percentage of layer digests of the image resulting from the original Dockerfile that also appear in the image built from the generated/retrieved Dockerfile. Note that if this measure is 100\%, in this context, it means that the generated/retrieved Dockerfile is able to provide everything that the original image already provided. Still, it is possible that it contains additional layers not present in the original image. 

\begin{table*}[t]
    \caption{Adherence score between the input and the generated \nlRecipe reported for each field. }
    \label{tab:rqOneScore}
    \centering
      \resizebox{\linewidth}{!}{
        \begin{tabular}{ l | l  l  l  l  l  l  l  l  l  l }
        \toprule
        \textbf{Approach}           & \textbf{OS}     & \textbf{Pkg Man.}& \textbf{Dep.}  & \textbf{ENV}  & \textbf{ARG}   & \textbf{LABEL} & \textbf{EXPOSE} & \textbf{CMD}  & \textbf{ENTRYPOINT}& \textbf{Down. Ext. Dep.}  \\
        \midrule
        \approach                   & \textbf{0.998}  & 0.981            & 0.865          & \textbf{0.892} & \textbf{0.987} & \textbf{0.999} & 0.798          & 0.743          & 0.843             & 0.816     \\
        \baseElastic                & 0.922           & \textbf{1.000}   & \textbf{0.877} & 0.812          & 0.884          & 0.872          & \textbf{0.829} & \textbf{0.826} & \textbf{0.851}    & \textbf{0.842}    \\
        \baseSentT                  & 0.880           & \textbf{1.000}   & 0.761          & 0.518          & 0.168          & 0.165          & 0.373          & 0.453          & 0.260             & 0.448    \\  
    
        \bottomrule
        \end{tabular}
      }
      \vspace{-0.5cm}
    \end{table*}  

\subsection{Data Availability}
We provide a replication package \cite{replication} containing the tool to extract high-level requirements from Dockerfiles, the trained DL models, along with the datasets and the code for replicating the baseline techniques and the conducted experiment.

\section{Empirical Study Results}
\label{sec:results}

\begin{figure}[t]
	\centering
	\includegraphics[width=\linewidth]{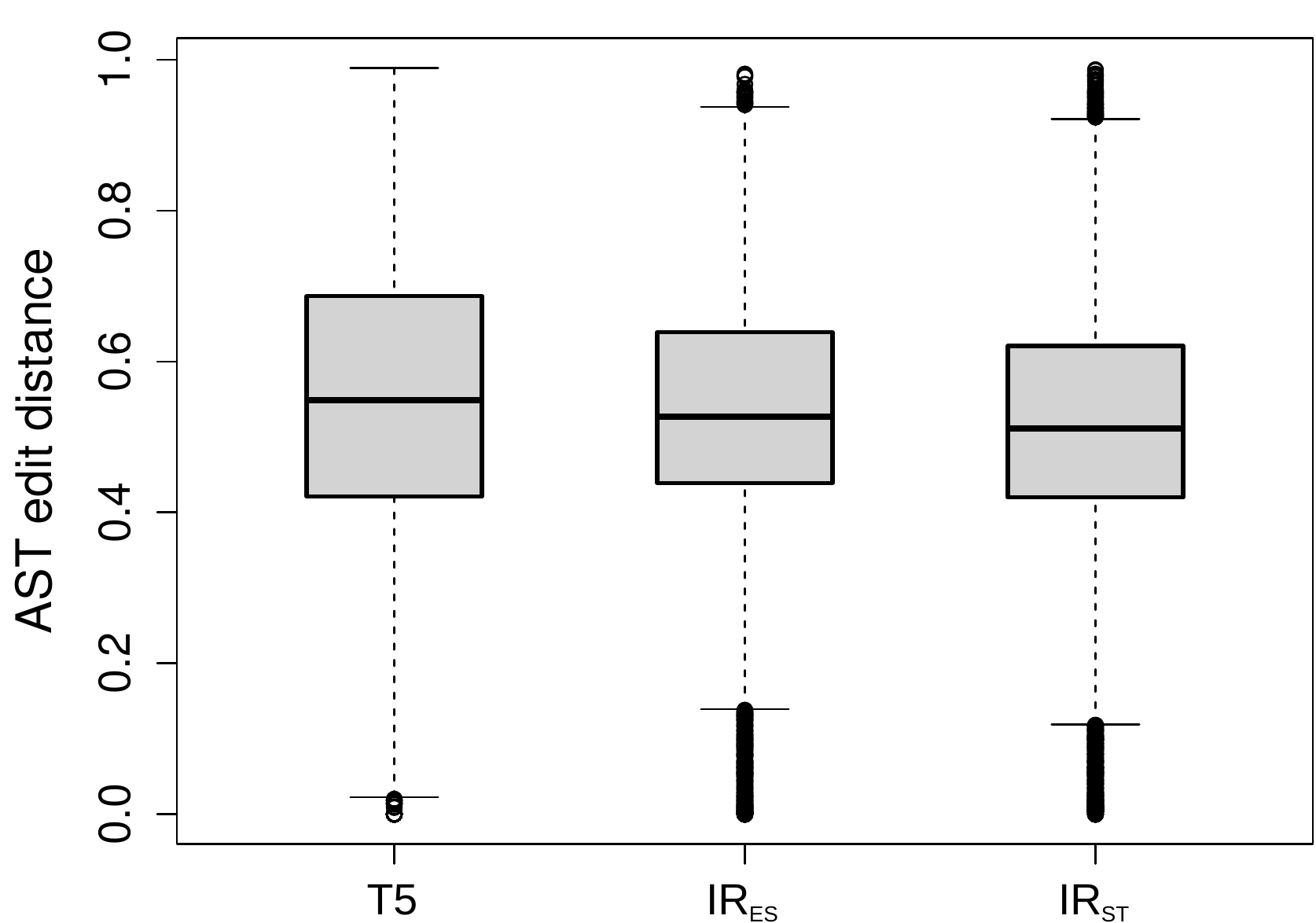}
	\caption{Boxplots of the normalized AST edit distance (\RQ{2}).}
	\label{fig:rqTwoBoxplot}
	\vspace{-0.3cm}
\end{figure}

In this section, we report the results of our study and, thus, the answers to our research questions.

\subsection{\RQ{1}: Adherence to the High-Level Specification}

In \tabref{tab:rqOneScore}, we report the results of the comparison with input \nlRecipe fields. 
First, it can be noticed that \baseElastic is the best-performing baseline, since it always achieves better or equal results than \baseSentT. Therefore, from now on, we only discuss the comparison between \approach and such a baseline in this RQ.
\approach is generally able to better meet the requirement in terms of \fieldOs (+7.6pp), while it achieves slightly worse results in terms of \fieldPkgMan (-1.9pp), \fieldReq (-1.2pp), and \fieldDownExtPkgs (-2.6pp). The last two are probably the most critical and hard-to-meet requirements since they also interact with each other, and we can observe that both the approaches generally achieve good results. As for the other requirements, we observe that \approach performs better on \fieldUsesEnv, \fieldUsesArg, and \fieldUsesLabel, while \baseElastic achieves better results in terms of \fieldUsesExpose, \fieldUsesCmd, and \fieldUsesEntrypoint. In summary, we can conclude that (i) there is no clear winner between the two approaches, and (ii) both the approaches generally return Dockerfiles that meet most of the requirements.

In summary, \approach and \baseElastic perform very similarly in terms of adherence to the input requirements: There is no clear winner as for this aspect.


\subsection{\RQ{2}: Dockerfile Similarity}
\begin{figure}[t]
	\centering
	\includegraphics[width=\linewidth]{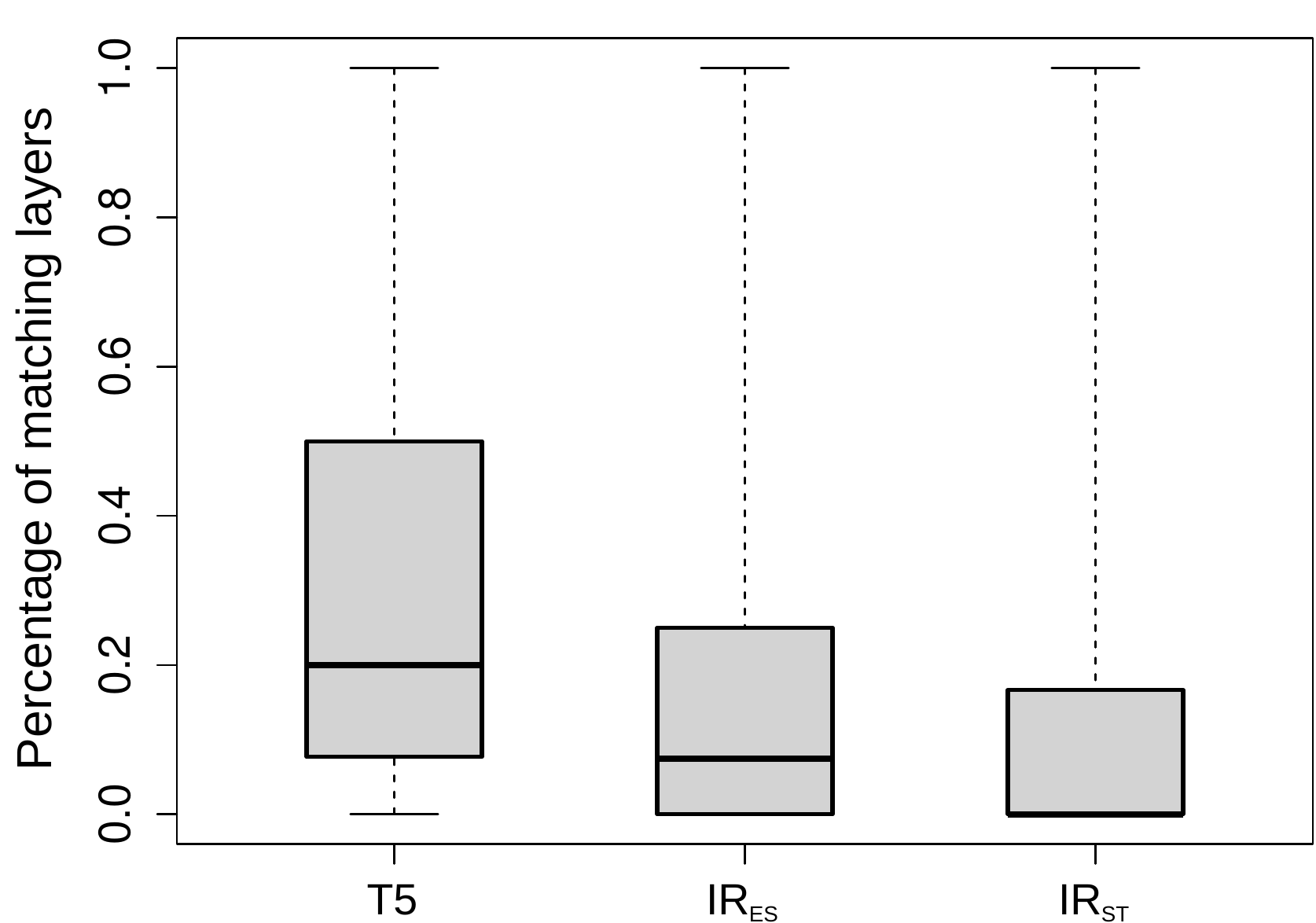}
	\caption{Boxplots of the percentage of matching layers (\RQ{3}).}
	\label{fig:rqThreeMatchingLayers}
	\vspace{-0.3cm}
\end{figure}

We report the adjusted boxplots \cite{hubert2008adjusted} for the normalized AST \textit{edit distance} in \figref{fig:rqTwoBoxplot}.
The boxplot shows the distribution of the edit distance between the generated Dockerfile and the original Dockerfiles for each instance of the test set. The higher the distance, the lower the similarity.
Also in this case, it seems that there is no clear winner: \approach has higher variance, thus being able to generate both  better and  worse Dockerfile compared to the two baselines. The mean edit distance is 0.55 ($\sigma$ = 0.19) for \approach, 0.53 ($\sigma$ = 0.18) for \baseElastic, and 0.51 ($\sigma$ = 0.19) for \baseSentT. This means that the two IR-based baselines perform slightly better than \approach. 
The difference is significant according to the Mann–Whitney U tests we performed for comparing \approach with the \baseElastic and \baseSentT (adjusted $p$-value lower than 0.001 for both).
The Cliff's Delta between \approach and \baseElastic is 0.06, and 0.11 between \approach and \baseSentT. Thus, the difference is \textit{negligible} in both cases.
We took a closer look at the cases in which the generated/retrieved Dockerfile was perfectly equal to the original one (\ie edit distance 0). We have 93 of such cases for \approach, while only 18 and 11 for \baseElastic and \baseSentT, respectively. In terms of AST size, the perfect matches are rather small (10.1, $\sigma$ = 9.4, with a maximum of 58) compared to the average size (133.7, $\sigma$ = 132.5) for \approach, while it is remarkably higher for the two baselines (39.5, $\sigma$ = 38.2 for \baseElastic, and 41.8, $\sigma$ = 44.8 for \baseSentT). Such a result suggests that \approach works well when small Dockerfile need to be generated, while it struggles with bigger ones. To confirm such a conjecture, we computed the correlation (Spearman $\rho$) between the AST size of the target Dockerfile for \approach, and we found that it is significant and high ($\rho$ = 0.67), much more than the two baselines ($\rho$ = -0.12 for \baseElastic, and not significantly different from 0 for \baseSentT).

In summary, all the approaches returns Dockerfiles quite different from the target ones. \approach works well with small Dockerfiles, but not with bigger ones.


\subsection{\RQ{3}: Docker Images Similarity}

\approach achieves a build success rate of 34\% (170/500 correctly built images), outperforming the \baseElastic (23\%, 166/500 images) and \baseSentT (32\%, 156/500 images).
Comparing the \textit{digest} of the built images (\ie hash value) with the source image (\ie the one built from the test instance), we obtain remarkably better results for \approach: We have 11.7\% of matches (20/170 instances), while the two baselines have no matching digest for their images. This result is confirmed also in \figref{fig:rqThreeMatchingLayers}, which depicts the distribution of the percentage of matching layers (adjusted boxplots \cite{hubert2008adjusted}). The mean percentage of matching layers is 32.4\% for \approach ($\sigma$ = 0.32), 17.5\% for \baseElastic ($\sigma$ = 0.26), and 10.9\% for \baseSentT ($\sigma$ = 0.19).
The obtained results complement the ones presented in \RQ{2}. Not only \approach works better when it needs to generate small Dockerfiles, but it also works better than the two baselines on bigger ones, up to a certain point; after that, it is not able to generate good instructions, thus the limited layer match, in absolute terms.

In summary, \approach achieves the best results compared to the two baselines in terms of build success, percentage of perfectly matching images, and percentage of matching layers.
\begin{figure}[t]
  \centering

  \begin{subfigure}{0.9\linewidth}
    \centering
\begin{lstlisting}
FROM golang:1.9.4-stretch

RUN apt-get update -y && apt-get upgrade -y 

RUN git clone https://github.com/edenhill/librdkafka.git && \
  cd librdkafka && \
  ./configure --prefix /usr && \
  make && \
  make install && \
  cd ..

WORKDIR /go/src/consumerpg
COPY . .

RUN go-wrapper download
RUN go-wrapper install

CMD ["go-wrapper", "run"]
\end{lstlisting}
    \vspace{-0.3cm}
    \caption{Target Dockerfile}
    \medskip
    \label{fig:target}
    \end{subfigure}
  
    \begin{subfigure}{0.9\linewidth}
        \centering
\begin{lstlisting}
FROM golang:1.9.4-stretch 

RUN apt-get update && apt-get install -y git 

RUN git clone https://github.com/edenhill/librdkafka.git && cd librdkafka && ./configure && make && make install 

RUN go get github.com/confluentinc/confluent-kafka-go 

WORKDIR /go/src/github.com/confluentinc/confluent-kafka-go 

CMD ["go", "run", "main.go
\end{lstlisting}
        \vspace{-0.3cm}
        \caption{Dockerfile generated by \approach}
        \label{fig:incompleteGen}
    \end{subfigure}
    \medskip
    \caption{Example of a generated incomplete Dockerfile}
  \vspace{-0.5cm}
\end{figure}


\section{Discussion}
\label{sec:discussion}




Generating Dockerfiles from high-level specifications is a challenging task.
In this study, we perform a first attempt to solve this problem using Deep Learning (\approach, specifically).
Considering the overall results, there is no clear evidence that \approach is better than using IR-based techniques, in practice. Given the lower effort in setting-up a IR-based technique (which is trivial in the case of \baseElastic, for example), at a superficial level, we can conclude that, at the moment, this would be the best option for practitioners.

However, we analyzed the results more in-depth to try to understand what went wrong, and why \approach does not work well for this task, despite it is works very well for other coding tasks \cite{mastropaolo2022tse}.
First, we observed that \approach generates Dockerfiles with a much lower number of tokens compared to the two baseline approaches (36.80 vs. 135.70 for \baseElastic and 107.10 for \baseSentT). This explains why \approach works well for smaller Dockerfiles and gradually less well for bigger ones (\RQ{2}), and also why \approach achieves a good percentage of matching layers even if the Dockerfile similarity is low (\RQ{3}).


We manually analyzed some Dockerfiles generated with \approach. We found that, in some instances, the Dockerfiles abruptly interrupt in the middle of the last instruction. While the remainder of the generated Dockerfiles is correct, the last instruction often contains issues. An example is provided in \figref{fig:incompleteGen}, with the target Dockerfile (a) and the one generated by \approach (b). The prediction is remarkably good, until, in the last line, \approach stops the generation at a certain point.
A pattern we observed is that interrupted Dockerfiles do not end with the token we used for indicating the new line (\texttt{<nl>}), while the ones in the training always end with such a token by design. We counted a total of 6,786 instances of such a kind ($\sim$60\% of the cases). In a real-world usage scenario, the developer would need to manually complete such Dockerfiles to make them work.

If we consider only the instances that terminate with the newline token, the results of \approach become better than the two IR baselines in all the aspects we considered: For \RQ{1}, \approach achieves better results than both the baselines for all the fields, except for \fieldPkgMan; for \RQ{2}, the edit distance becomes significantly lower, \ie 0.46, compared to 0.51 of the best baseline for such a sub-sample; For \RQ{3}, we obtain results in line with the previously presented ones.
We tried to address this issue by replacing the greedy decoding strategy with a sampling strategy, with different values for the \textit{temperature} hyper-parameter of the \textit{softmax} function. A high \textit{temperature} allows to increase the chances of picking tokens with lower likelihood, while a lower \textit{temperature} does the exact opposite, so that, when the temperature is close to 0, such a decoding strategy behaves like a greedy decoding strategy. 

We tested temperature values between 0.7 and 1.0, with a step of 0.1.
We observe that increasing the \textit{temperature} allows to reduce the number of incomplete Dockerfiles: With a \textit{temperature} of 0.7, we obtain a total of 1,648 incomplete generations (14.6\%) which decrease to 1,283 (11.3\%) with a \textit{temperature} of 1.
The average number of tokens contained in the Dockerfiles varies between 82 (\textit{temperature} = 1.0) and 84 (\textit{temperature} = 0.7), while with greedy decoding we have 37, on average.
In the end, however, the Dockerfiles generated with this strategy achieve generally worse results in terms of (i) number perfect predictions ($\sim$-28\%, with the best \textit{temperature}, \ie 0.8), (ii) number of perfectly matching images (-$\sim$80\%, best \textit{temperature} = 1.0) and layers (-$\sim$20\%, best \textit{temperature} = 0.9). 

This analysis shows that, while \approach learned how to generate Dockerfiles, to some extent, it does not have enough knowledge to  generate complete Dockerfiles well: It either partially generates good parts of Dockerfiles or generates complete less-good Dockerfiles.
There are two possible explanations for this phenomenon, which also represent open problems for this specific task:
\begin{itemize}
  \item \textbf{A larger dataset needs to be built.} Addressing this problem is only apparently easy. We considered in our study the largest collection of Dockerfiles available in the literature \cite{eng2021revisiting}, which includes \textit{all} the Dockerfiles from open-source projects produced up to 2020. While such a dataset can be updated with the last two years of activities in GitHub, it is  unlikely that the size of our dataset would drastically increase as a result. Indeed, we consider a single Dockerfile for each unique \nlRecipe, \ie we would not have new instances for the \nlRecipes already covered. We rely on comments by the authors to extract some requirements (specifically, the \fieldReq field). Some Dockerfiles, however, do not have explicit indications of such requirements and, therefore, we miss them in the inferred specifications. New and more precise ways of extracting high-level requirements from Dockerfiles are needed. Also, a promising direction would consist in the definition of techniques for data augmentation in this context, \eg by blending existing Dockerfiles to provide uncovered combinations of \nlRecipes.
   
  \item \textbf{A different training stop criterion needs to be defined.} In this study, we used the same procedure previously used for coding tasks \cite{mastropaolo2022tse}, for which the problem of abrupt interruption in the inference has not been observed. It is possible that the stop criterion and the metric (\textit{BLEU}) used are not the right ones in this context. As for the latter, it might be worth exploring different distance measures. While the AST distance is not a viable option for performance reasons, other metrics that do not consider the order in the instructions might be more useful.
\end{itemize}

As a takeaway for practitioners, it may be too early to reliably use these approaches to generate entire Dockerfiles in one shot, without the need to apply minor adjustments. 


\section{Threats to Validity}

\textit{Threats to Construct Validity} concern the correct operationalization of the concepts being studied. First, the inference method we used for extracting requirements from Dockerfiles works on a series of assumptions (\eg the presence of comments) that might not always be completely satisfied in practice. To mitigate this crucial threat, we carefully tested our parser and manually checked examples until we were satisfied with the procedure used to achieve this goal. In total, we were able to infer 113,442 unique \nlRecipes, with 31,990 unique combination of dependencies, which gives us confidence on the fact that our parser works as intended for most of the Dockerfiles considered. Also, we made sure to exploit all the instances (also the ones without comments) in the training procedure (\ie by using them for pre-training).
The choice of the fields that compose our \nlRecipe could exclude requirements that developers might be interested in specifying (such as \texttt{USER}). As we explained in \secref{sec:approach}, we exclude only the requirements derived from Dockerfile instructions that do not appear very often.
The best candidate Dockerfile we selected for each \nlRecipe (\secref{sec:approach}) might not be representative of the group of Dockerfiles. In our methodology, we relied on Jaccard similarity to mitigate this risk.

\textit{Threats to Internal Validity} concern factors internal to our study that might have influenced our findings.
The percentage of marching layers (\RQ{3}) might not accurately capture the structural similarity between two Docker images: If a layer is different, many or even all the subsequent layers will be different.
However, to the best of our knowledge, the only alternative is to perform a diff on Docker containers \cite{web:containerdiff}, which, however, only works at the level of installed packages.
Considering the layers allows knowing from which point the two images started to differ, we believe this is the best choice.

\textit{Threats to External Validity} concern the generalizability of our findings.
We relied on the largest collection of Dockerfiles available in the literature \cite{eng2021revisiting}. More Dockerfiles might have been created between 2020 and now. However, we believe that the considered dataset allows us to provide reasonably generalizable.

\section{Conclusion}
We evaluated the effectiveness of Deep Learning (and, specifically, \approach) for the automatic generation of Dockerfiles. 
The results show that, while \approach works very well on small Dockerfiles, it struggles with larger ones.
After having deeply analyzed this phenomenon, we identified two possible issues that must be addressed before deploying a working DL-based solution for this task. First, it is necessary to \textit{build a larger dataset}, which is not easy, given that we used the largest collection of open-source Dockerfiles available. Second, it is necessary to \textit{devise a different stopping criterion} for fine-tuning \approach since the one typically used for coding tasks (based on BLEU-4) likely causes an early stop, which does not allow the model to properly complete the learning process.



\bibliography{main}
\bibliographystyle{IEEEtran}

\end{document}